# Sub-barrier heavy-ion fusion/capture: Accurate accounting for zero-point quadrupole shape oscillations with realistic nucleus-nucleus potential

I. I. Gontchar[1*], M. V. Chushnyakova[2]

[1]Physics and Chemistry Department, Omsk State Transport University, Omsk 644046, Russia
[2]Physics Department, Omsk State Technical University, Omsk 644050, Russia

* Corresponding author. E-mail: vigichar@hotmail.com

**Abstract** Although the study of collisions of heavy ions resulting in formation of dinuclear systems has a long history this process still is a subject of experimental and scientific activities, partly because nowadays the heavy-ion fusion process is the only practical way for extending further the Periodic Table. Yet the heavy-ion capture cross sections are calculated theoretically with significant uncertainties. In particular, these cross sections below the barrier sometimes underestimate the experimental ones by orders of magnitude. In the literature, it had been shown that accounting for the Zero-Point Oscillations (ZPO) of the shape of colliding nuclei increased significantly the calculated capture cross sections. However, those calculations had been performed in a simplified way with a schematic nucleus-nucleus potential and for few reactions. The purpose of the present paper is to account for the ZPO in a more realistic way and to compare systematically the resulting capture cross sections with the experimental data. In the present paper, the nucleus-nucleus potential is evaluated using the semi-microscopical double-folding model with M3Y-Paris nucleon-nucleon forces. The nucleon densities needed for finding this potential are generated from the experimental three-parameter Fermi charge densities. The calculated capture cross sections are compared with the data for reactions $^{16}O + {}^{54}Fe, {}^{58}Ni, {}^{144}Nd, {}^{148}Sm$ and $^{58}Ni + {}^{54}Fe, {}^{58}Ni, {}^{60}Ni$. We show that there are certain regularities in the relationship between the experimental and theoretical cross sections with respect to the period of the quadrupole vibrations of the colliding nuclei at their ground states.

## 1. Introduction

Capture in the orbital motion is the first step for the processes resulting from collision of two complex nuclei ("heavy ions") at the energies around Coulomb barrier [1-5]. The capture can be followed either by fusion with formation of compound nucleus (which after either fissions or survives as an evaporation residue) or by fast fission (quasifission) when the compact compound nucleus configuration is not reached, and collision partners are reseparated with the asymmetry significantly different than the initial one [6-11]. Therefore, the capture and fusion cross sections, $\sigma_{cap}$ and $\sigma_{fus}$, are in general rather different: $\sigma_{cap} \geq \sigma_{fus}$. However, for relatively light reactions these two cross sections are very close. The border value of the product of the charge numbers of projectile ($Z_P$) and target ($Z_T$) nuclei was found theoretically to be about 1600 in [8] and later experimentally about 800 in [1].

A long time ago, it had been demonstrated [12-14] that accounting for the zero-point oscillations (ZPO) of the nuclear shapes significantly improved the agreement of the calculated capture cross sections in heavy-ion collisions with the data. In the present work, we attempt to elaborate and modify the approach of [12]. The proposed improvements are the following:
    i) the strong nucleus-nucleus potential (SnnP) is evaluated using the semi-microscopic double folding model (DFM) with the M3Y-Paris nucleon-nucleon forces;
    ii) the zero-point oscillations are accounted for in a more detailed way and for both collision partners.

Note that, in some recent works, the M3Y-Reid forces [15] are used [16-18] despite it is stated in [19] that "the Paris potential is based on a more fundamental theory of NN-scattering than the earlier potentials... The Reid soft-core potential is based on earlier and partially erroneous phase-shift data..."

Comparing the calculated capture cross sections with the experimental ones we will use seven reactions for which $Z_P Z_T < 800$ (see Table 3 below), i.e. those for which $\sigma_{cap} = \sigma_{fus}$. That is why henceforth we simply use the term "cross sections" or the abbreviation CSs. Note that even for such reactions the theoretical description is a subject of some uncertainties.

In many experimental works (see e.g. [20-25]) the theoretical analysis of the measured CSs often is performed using the coupled-channels method [26-29]. The bare nucleus-nucleus potential in this analysis is taken typically to have the Woods-Saxon shape. The parameters of this potential are fitted to reproduce the measured above-barrier cross sections individually for each reaction. The diffuseness of this Woods-Saxon potential was shown to exceed significantly the one extracted from the analysis of elastic scattering data [21, 22, 30, 31]. In our opinion, these circumstances represent serious drawbacks of the coupled-channels method. We believe that the title of the widely known article [30] "Systematic failure of the Woods-Saxon nuclear potential to describe both fusion and elastic scattering: Possible need for a new dynamical approach to fusion" indicates the same.

On the other hand, in many works [32-39] the above-barrier CSs had been described within the framework of the models accounting for dissipation. That is why we expect that our present calculations ignoring dissipation should overestimate the data above the barrier. Below the barrier we hope to find the same situation as in [12].

The paper is organized as follows. Section 2 is devoted to a brief description of the SnnP calculation routine. The choice of the nuclei and reactions for the analysis is explained in Section 3. The resulting nucleus-nucleus potentials are discussed in Section 4. Section 5 contains the validation of the developed model and main results for the cross sections in comparison with the experimental data. Our conclusions are formulated in Section 6.

## 2. Nucleus-nucleus potential

In the DFM with the M3Y nucleon-nucleon (NN) forces approach, the nucleus-nucleus potential $U$ reads

$$U(R, \theta_P, \theta_T) = U_C(R, \theta_P, \theta_T) + U_{nD}(R, \theta_P, \theta_T) + U_{nE}(R, \theta_P, \theta_T). \tag{1}$$

Here the Coulomb term $U_C$, the direct $U_{nD}$ and exchange $U_{nE}$ parts of SnnP $U_n$ depend upon the distance between the centers of mass of projectile (P) and target (T) nuclei and their orientation angles $\theta_P, \theta_T$ (see Fig. 1). The three terms of Eq. (1) read

$$U_C = \int d\vec{r}_P \int d\vec{r}_T \rho_{qP}(\vec{r}_P) v_C(s) \rho_{qT}(\vec{r}_T), \tag{2}$$

$$U_{nD} = \int d\vec{r}_P \int d\vec{r}_T \rho_{AP}(\vec{r}_P) v_D(s) \rho_{AT}(\vec{r}_T), \tag{3}$$

$$U_{nE} = \int d\vec{r}_P \int d\vec{r}_T \rho_{AP}(\vec{r}_P) v_E(s) \rho_{AT}(\vec{r}_T). \tag{4}$$

Here $\rho_{AP}$ and $\rho_{AT}$ ($\rho_{qP}$ and $\rho_{qT}$) stand for the nucleon (charge) densities, $\vec{r}_P$ and $\vec{r}_T$ are the radius-vectors of the interacting points of the projectile and target nuclei,

$$\vec{s} = \vec{R} + \vec{r}_T - \vec{r}_P \tag{5}$$

(see Fig. 1).

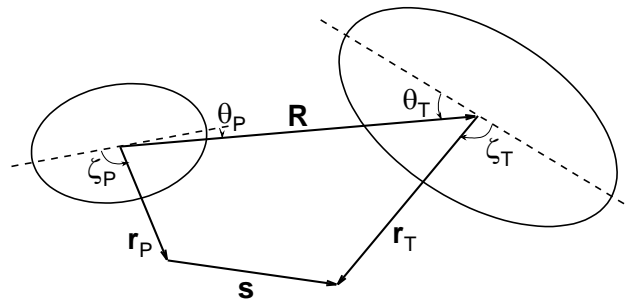

**Fig. 1.** Schematic illustration of the collision geometry (see text for explanations).

In Eqs. (2)-(4), one neglects possible time-dependence of the densities (the so-called frozen density approximation). This approximation seems to work reasonably well unless the density overlap of the colliding nuclei is about 1/3 of the saturation density 0.16 fm$^{-3}$ (see, e.g. [40-43]). In other words, the barrier radius should, at least, exceed the sum of the nuclear radii. This ratio is presented in Table 3 below for all reactions studied corroborating the validity of the frozen density approximation for our present study.

Let us note that the density-dependent M3Y *NN*-forces are considered in the literature only for spherical colliding nuclei. It is not clear how to account for the density dependence when the nuclei are deformed. That is why we ignore this dependence in Eqs. (3)-(4).

The direct part of the effective *NN*-interaction $v_D(s)$ consists of two Yukawa terms [16, 17, 19, 44]:

$$v_D(s) = \sum_{i=1}^{2} G_{Di} \exp(-s/r_{vi})/(s/r_{vi}). \tag{6}$$

For the exchange part $v_E(s)$, one finds in the literature two options: an advanced and complicated one with the finite range (in this case, Eq. (4) becomes much more complicated) and a simpler one with zero range [17, 43, 45, 46]. The latter version reads:

$$v_E(s) = G_{E\delta}\, \delta(\vec{s}). \tag{7}$$

In [46] it has been shown that varying the value of $G_{E\delta}$ with respect to its standard value $-592$ MeV fm$^3$ from Ref. [19] down to $-1040$ MeV fm$^3$ makes it possible to reproduce the Coulomb barrier energies resulting from the option with the finite range exchange force. Note that the computer calculations with the zero-range exchange forces run significantly (by a factor of 100) faster than the ones with the finite range option. These are Eqs. (6),(7) which are employed in the present paper (the values of the coefficients are collected in Table 1). The real calculations are performed using the Fourier transform (momentum space representation). The details can be found in [47, 48].

**Table 1**
The parameters of the Paris M3Y interactions used in Eqs. (6),(7).

| Parameter | $G_{D1}$ (MeV) | $G_{D2}$ (MeV) | $G_{E\delta}$ (MeV fm$^3$) | $r_{v1}$ (fm) | $r_{v2}$ (fm) |
|---|---|---|---|---|---|
| Value | 11062 | $-2538$ | $-1040$ | 0.25 | 0.40 |

For the nucleon and charge densities in Eqs. (2)-(4), we use the three-parameter Fermi formula (3pF-formula):

$$\rho_F(r) = \rho_{CF} \frac{1 - w_F r^2/R_F^2}{1 + \exp[(r - R_F)/a_F]}. \tag{8}$$

In Eq. (8), $R_F$ corresponds approximately to the half central density radius, $a_F$ is the diffuseness, $\rho_{CF}$ is defined by the normalization condition. The 3pF-formula (or its version with $w_F = 0$ called 2pF-formula) is often applied to approximate the experimental nuclear charge densities (see Ref. [49] and references therein). To consider ZPO, we allow the spherical nuclei to become deformed. In this case, one should account for the density dependence upon $\zeta$:

$$\rho_F(r, \zeta) = \rho_{CF} \frac{1 - w_F r^2/[R_F f(\zeta)]^2}{1 + \exp\{[r - R_F f(\zeta)]/a_F\}} \tag{9}$$

where

$$f(\zeta) = \lambda^{-1}\bigl(1 + \beta_2 Y_{20}(\zeta)\bigr). \tag{10}$$

In Eq. (10), $\lambda$ is responsible for the volume conservation, $Y_{20}$ is the spherical function, $\beta_2$ stands for the quadrupole deformation parameter. For the meaning of $\zeta$, see Fig. 1. Since we do not consider other deformations, we omit the subscript "2" below, i.e. $\beta \equiv \beta_2$.

We use the same 3pF-formulas for the proton, neutron, and charge densities of a given nucleus. The parameters $R_{Fq}$, $a_{Fq}$, and $w_{Fq}$ of the charge density are taken from [49]. The half-density radii for proton $R_{Fp}$ and neutron $R_{Fn}$ densities are taken to be equal to $R_{Fq}$; the same is true for $w_F$. The proton $a_{Fp}$ and neutron $a_{Fn}$ diffusenesses are calculated approximately via the charge diffuseness $a_{Fq}$ [40, 50]:

$$a_{Fp} = a_{Fn} = \sqrt{a_{Fq}^2 - \frac{5}{7\pi^2}\left(0.76 - 0.11\frac{N}{Z}\right)}. \tag{11}$$

## 3. Choosing nuclei and reactions

We have selected the nuclei and reactions for the present study from the following considerations. First, we use only those spherical nuclei for which in [49] the 3pF-formulas are found for the experimental charge densities. Second, we select the reactions for which the experimental data on the CSs are available for the collision energy $E_{c.m.}$ covering the barrier region. These happen to be serious limitations: we have found only 9 reactions for the study. For instance, nice capture data are available for $^{16}O + ^{92}Zr$ [21] and $^{16}O + ^{144}Sm$ [51] but for $^{92}Zr$ and $^{144}Sm$ the 3pF-approximation for the charge distributions is absent in [49].

Many data for the cross sections are adopted from the database [52]. Information on the projectile and target nuclei is comprised in Table 2; information on the reactions chosen for consideration is presented in Table 3.

**Table 2**
Parameters of the nuclei involved in the analysis: charge density parameters $R_{Fq}$, $a_{Fq}$, $w_{Fq}$ of Eqs. (8), (9) taken from [49]; $E2+$, $B(E2\uparrow)$, $S = \sqrt{\langle\beta^2\rangle}$ taken from the database [52]; the period of quadrupole oscillations calculated as $\tau_0 = 2\pi\hbar/E2+$.

|  | $^{16}O$ | $^{54}Fe$ | $^{58}Ni$ | $^{60}Ni$ | $^{144}Nd$ | $^{148}Sm$ |
|---|---|---|---|---|---|---|
| $R_{Fq}$ (fm) | 2.608 | 4.106 | 4.309 | 4.489 | 5.6256 | 5.771 |
| $a_{Fq}$ (fm) | 0.513 | 0.519 | 0.517 | 0.537 | 0.6178 | 0.596 |
| $w_{Fq}$ | −0.051 | 0.000 | −0.131 | −0.267 | 0.000 | 0.000 |
| $E2+$ (MeV) | 6.917 | 1.408 | 1.454 | 1.333 | 0.696 | 0.550 |
| $B(E2\uparrow)$, ($e^2b^2$) | 0.00371 | 0.0608 | 0.0650 | 0.0916 | 0.504 | 0.713 |
| $S$ |  | 0.349 | 0.193 | 0.177 | 0.205 | 0.125 | 0.142 |
| $\tau_0$ (zs) | 0.598 | 2.935 | 2.840 | 3.102 | 5.933 | 7.509 |

The value $S$ is called "zero-point amplitude" [53] or "amplitude of zero-point motion" [54] and is defined as

$$S^2 = \langle\beta^2\rangle_0 = \frac{\hbar\omega_\beta}{2C_\beta}. \tag{12}$$

Subscript '0' indicates that the averaging is performed at the ground state. In Eq. (12), there may stay another collective variable describing quadrupole deformation and proportional to $\beta$; in this case, the stiffness $C_\beta$ must be changed to an appropriate stiffness (the frequency $\omega_\beta$ remains unchanged).

The zero-point amplitude is related to $B(E2\uparrow)$ as [42]

$$S = \frac{4\pi}{3r_0^2 Z A^{2/3}} \sqrt{\frac{B(E2\uparrow)}{e^2}}. \tag{13}$$

Note that the values $\beta_2^0$ presented in the NRV database [52] for even-even nuclei are just the zero-point amplitude, i.e. $S = \beta_2^0$.

**Table 3**
The reactions under consideration with their ordering numbers which are used henceforth; Coulomb barrier energy at zero angular momentum for spherical reagents $B_{0sph}$; the ratio of the spherical barrier radius to the sum of spherical reagents half-density radii; reference to the experimental cross sections.

|    | Reaction | $Z_P Z_T$ | $B_{0sph}$ (MeV) | $\frac{R_{B0sph}}{R_{FqP} + R_{FqT}}$ | Source of $\sigma_{exp}$ |
|----|----------|-----------|------------------|---------------------------------------|--------------------------|
| R1 | $^{16}O + ^{58}Ni$ | 224 | 31.24 | 1.40 | [55] |
| R2 | $^{16}O + ^{54}Fe$ | 208 | 29.33 | 1.43 | [56] |
| R3 | $^{16}O + ^{144}Nd$ | 480 | 57.66 | 1.37 | [57, 58] |
| R4 | $^{16}O + ^{148}Sm$ | 496 | 60.04 | 1.34 | [51] |
| R5 | $^{58}Ni + ^{58}Ni$ | 784 | 98.00 | 1.24 | [59] |
| R6 | $^{58}Ni + ^{54}Fe$ | 728 | 92.79 | 1.26 | [60] |
| R7 | $^{58}Ni + ^{60}Ni$ | 784 | 97.40 | 1.23 | [61] |

## 4. Results: Nucleus-nucleus potential

The calculated CSs are defined principally by the potential energy profile. That is why we show these profiles for reaction R4 $^{16}O + ^{148}Sm$ in Fig. 2. The $U(R)$-dependencies are presented for the nose-to-nose geometry for three values of the angular momentum $J$ and three sets of deformations. These deformations come from quantum fluctuations of the quadrupole deformation parameters of projectile and target nuclei, $\beta_P$ and $\beta_T$, having the Gaussian distributions with zero average and rms $S_P$ and $S_T$ whose experimental values are tabulated in [52] and Table 2. The highest Coulomb barrier corresponds to the collision of two oblate nuclei. From Eq. (15) below, one sees that the higher the Coulomb barrier the smaller the transmission coefficient and consequently the smaller the cross section. Therefore, the collision of two oblate nuclei does not contribute significantly to fusion at sub-barrier collision energies. In Figs. 2 and 3, the highest Coulomb barrier corresponds to the case of two spherical nuclei; this version of calculations is noted as "ss" in contrast to two deformed colliding partners ("dd") and collision of spherical projectile and deformed target nuclei ("sd").

In each panel of Fig. 2, one sees that as the target nucleus becomes prolate the barrier decreases; as, in addition, the projectile nucleus becomes prolate the barrier decreases even more. The increase of angular momentum makes the barrier higher. These two observations are in agreement with the literature [47, 62-66]. It is probably more interesting to notice that the barriers are highly asymmetric, and it does not look reasonable to approximate them by parabolas at the level of $E_{c.m.}/B_{0sph} < 0.95$ ($B_{0sph}$ is the Coulomb barrier energy at zero angular momentum for spherical reagents). To corroborate this statement, we compare the transmission coefficients calculated using the WKB-approximation

$$T_J(\beta_P, \beta_T) = \left\{ 1 + \exp\left[\frac{2\Lambda(\beta_P, \beta_T)}{\hbar}\right] \right\}^{-1} \tag{14}$$

and calculated in the parabolic barrier approximation

$$T_J(\beta_P, \beta_T) = \left\{ 1 + \exp\left[\frac{2\pi(B - E_{c.m.})}{\hbar \omega_B}\right] \right\}^{-1}. \tag{15}$$

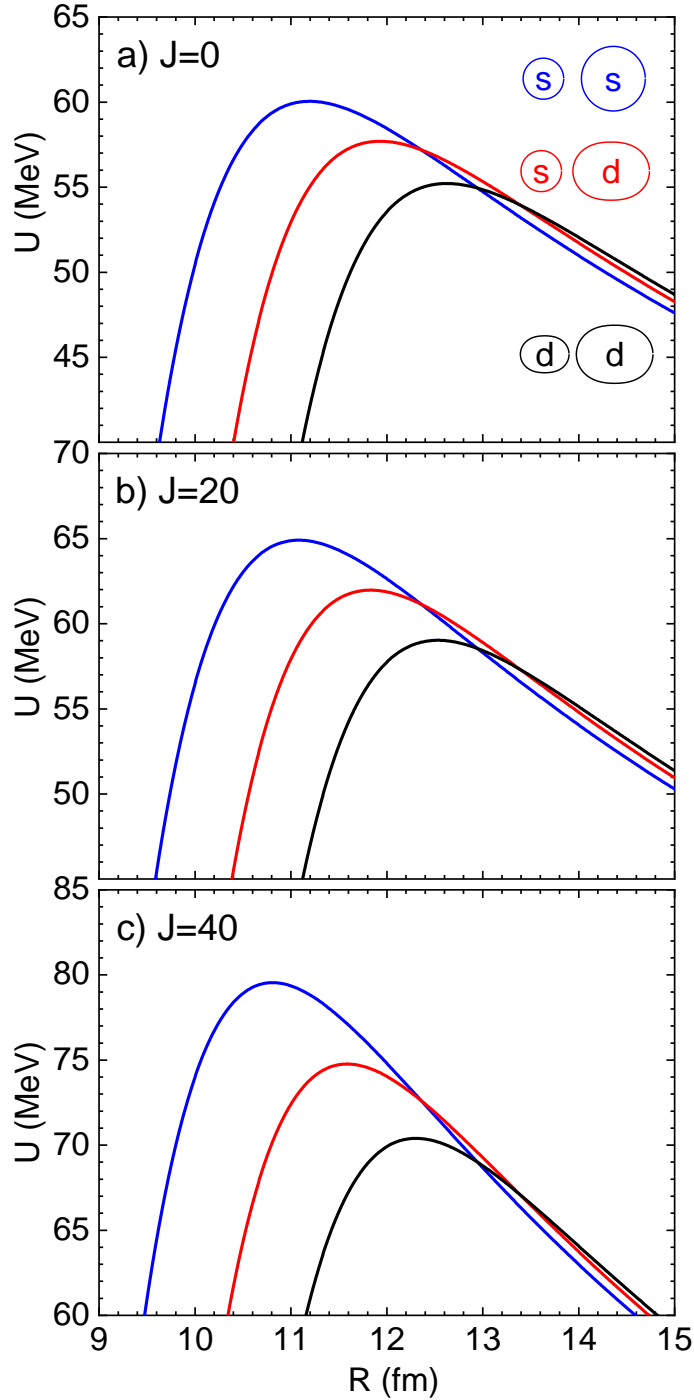

**Fig. 2.** Nucleus-nucleus potential versus $R$ at three values of $J$ in units of $\hbar$: (a) $J = 0$; (b) $J = 20$; (c) $J = 40$. In each panel, $U$ is presented for three combinations of deformations: the potential with the highest barrier corresponds to two spherical nuclei ("ss"); the one with the lowest barrier is for two equally deformed prolate nuclei (the largest deformation defined by Eqs. (22) below, "dd"); the intermediate potential corresponds to the collision of spherical projectile and deformed prolate target nuclei ("sd"). Calculations are performed for reaction R4 with $M = 13$, $k = 3.0$ (see Eqs. (22)). The shapes of colliding partners corresponding to the potentials are schematically shown in panel (a).

In Eq. (14), $\Lambda$ denotes the action calculated for the given angular momentum $J$ and collision energy $E_{c.m.}$ from the outer turning point down to the inner one. In Eq. (15), both the barrier energy $B$ and frequency $\omega_B$ are $J$, $\beta_P, \beta_T$-dependent. The transmission coefficients are compared in Fig. 3 for the reaction R4 (see Table 3) and the same three sets of deformations as in Fig. 2. The collision energy here is 95% of $B_{0sph}$. The difference between $T_J$ calculated by these two approaches at high values of $J$ exceeds a factor of 10. Only at $E_{c.m.} = B$ (i.e. when both reagents are deformed at $J = 0$ in panel c), the two approaches produce equal transmission coefficients. Thus, numerous calculations of the cross sections using the parabolic barrier approximation met in the literature should be taken with care.

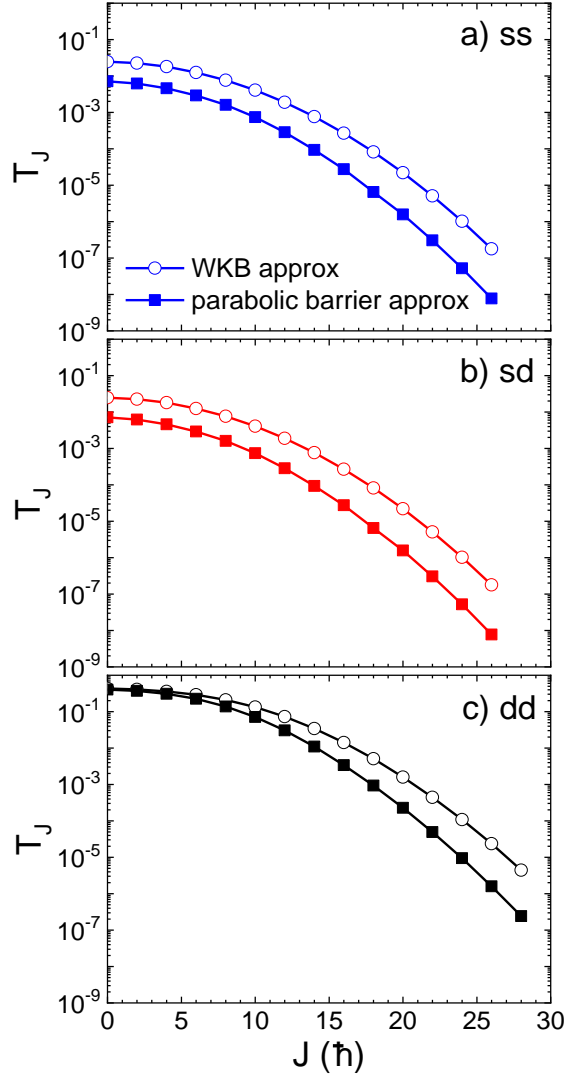

**Fig. 3.** The transmission coefficients versus $J$ for the same three combinations of deformations as in Fig. 2. Open circles correspond to the WKB calculations (Eq. (14)); closed squares stand for the parabolic barrier approximation (Eq. (15)). Calculations have been performed for reaction R4 with $M = 13, k = 3.0$ (see Eqs. (22)), $E_{c.m.} = 57.5$ MeV.

## 5. Results: Cross sections

We evaluate the cross section for non-deformed reagents by means of the standard formula

$$\sigma_{sph} = \frac{\pi \hbar^2}{2 m_R E_{c.m.}} \sum_J (2J + 1)\, T_J(\beta_P = 0, \beta_T = 0). \tag{16}$$

For each pair of deformations $\beta_P, \beta_T$ the cross section is evaluated as

$$\sigma_{\beta_P \beta_T} = \frac{\pi \hbar^2}{2 m_R E_{c.m.}} \sum_J (2J + 1)\, T_J(\beta_P, \beta_T). \tag{17}$$

Here $J$ is the angular momentum in units of $\hbar$, $m_R$ is the reduced mass. The summation over $J$ is terminated when the partial CS becomes $10^{-4}$ of its maximum value at fixed values of $\beta_P$ and $\beta_T$. The transmission coefficients are calculated using the WKB-approximation below the barrier and using the parabolic barrier approximation above the barrier. The WKB-approximation is widely used in modern studies [67-70] Note that we consider only nose-to-nose configurations following Refs. [32, 33].

The CS which is to be compared with the data is evaluated as follows:

$$\sigma_{th} = \sum_{i,j} \sigma_{\beta_{Pi} \beta_{Tj}} \, \Pi_P(\beta_{Pi}) \, \Pi_T(\beta_{Tj}). \tag{18}$$

Here $\Pi_P(\beta_P)$ and $\Pi_T(\beta_T)$ denote the probabilities for the projectile and target nuclei to possess the deformations $\beta_P$ and $\beta_T$ corresponding to the probability density for the harmonic oscillator at the ground state:

$$\Pi_P(\beta_{Pi}) = \Pi_{Pi} = N_P^{-1} \exp\left(-\frac{\beta_{Pi}^2}{2S_P^2}\right), \tag{19}$$

$$\Pi_T(\beta_{Tj}) = \Pi_{Tj} = N_T^{-1} \exp\left(-\frac{\beta_{Tj}^2}{2S_T^2}\right). \tag{20}$$

The model of harmonic oscillator is a standard one for the description of low-lying excited states of spherical nuclei. At the ground state the coordinate of this oscillator, i.e. the deformation parameter, possesses the Gaussian distribution (19), (20). The normalization factors $N_P$ and $N_T$ read

$$N_{P(T)} = \sum_{f=0}^{M-1} \exp\left(-\frac{\beta_{P(T)f}^2}{2S_{P(T)}^2}\right). \tag{21}$$

The sets of $M$ values of $\beta_{Pi}$ and $M$ values of $\beta_{Tj}$ are generated as follows:

$$\beta_{Pi} = kS_P \left(\frac{2i}{M-1} - 1\right), \quad \beta_{Tj} = kS_T \left(\frac{2j}{M-1} - 1\right). \tag{22}$$

Thus, we have two numerical parameters, $k$ and $M = 3, 5, 7, \ldots$ The resulting cross section $\sigma_{th}$ must become independent of both $k$ and $M$ when they are large enough. This validation is illustrated by Figs. 4 and 5.

In Fig. 4a, the CSs calculated using different $M = 5, 7, 9, 11, 13, 15$ are shown in linear and logarithmic scales versus $E_{c.m.}/B_{0sph}$. In such a presentation, the influence of $M$ cannot be seen clearly. Therefore, we show in Fig. 4b the fractional difference

$$\xi_M = \frac{\sigma_{th}(M)}{\sigma_{th}(M=15)} - 1. \tag{23}$$

One sees that $\sigma_{th}$ is sensitive to $M$ only below the barrier. With the accuracy of 10%, already $M = 5$ is suitable. As $M$ increases, $\xi_M$ goes to zero confirming the convergency of the calculated cross sections with respect to $M$.

The sensitivity of $\sigma_{th}$ to the value of parameter $k$ is illustrated by Fig. 5. Again, this sensitivity is significant only for $E_{c.m.} < B_{0sph}$. With the accuracy of 10%, $k = 2.6$ is acceptable. The values $M = 13$ and $k = 2.8$ are used for all the calculations below.

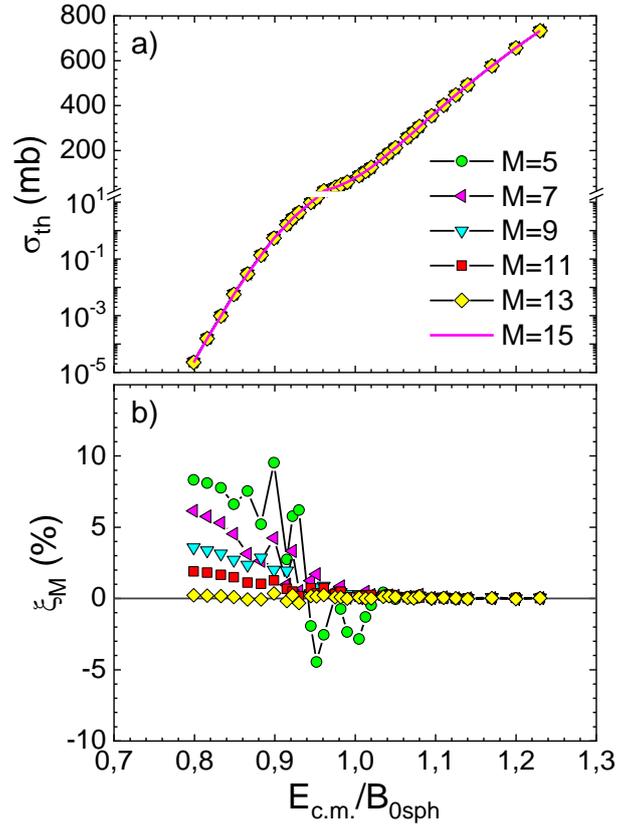

**Fig. 4.** Convergency of the calculated cross sections $\sigma_{th}$ with respect to parameter $M$ of Eqs. (22): (a) the CSs calculated at 6 values of $M$ indicated in the panel in linear and logarithmic scales; (b) the fractional difference $\xi_M$ (see Eq. (23)). Reaction R4; $B_{0sph} = 60.0$ MeV.

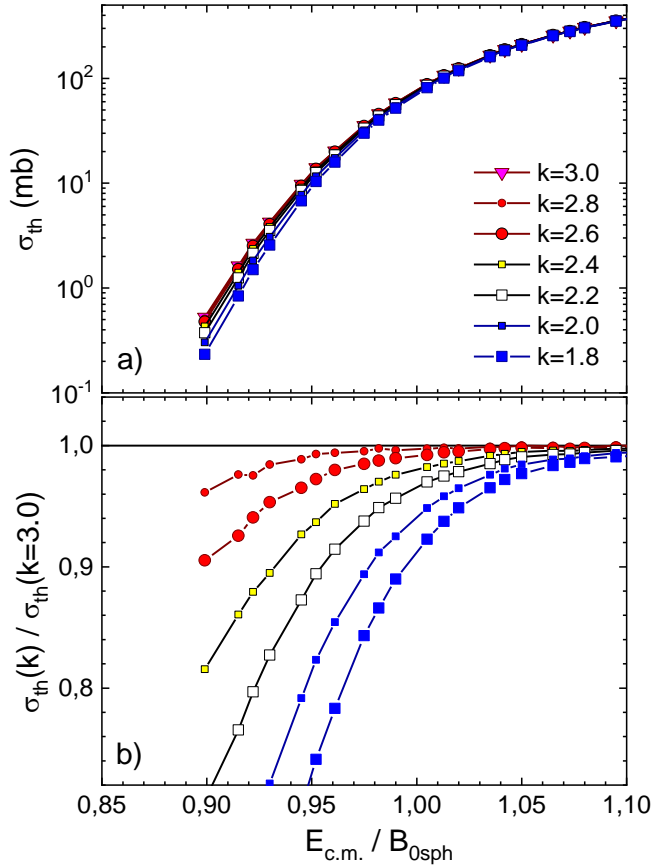

**Fig. 5.** Convergency of the calculated cross sections $\sigma_{th}$ with respect to parameter $k$ of Eqs. (22): (a) the CSs calculated at 7 values of parameter $k$ indicated in the panel; (b) the ratio $\sigma_{th}(k)/\sigma_{th}(k=3.0)$. Reaction R4; $B_{0sph} = 60.0$ MeV.

In Figs. 6,7 we compare the calculated cross sections $\sigma_{th}$ and $\sigma_{sph}$ with the experimental ones $\sigma_{exp}$. All reactions are split into two groups: the first group (R1-R4) involving $^{16}$O; the second one (R5-R7) involving $^{58}$Ni. In each group the reactions are arranged in increasing order of the period of quadrupole oscillations $\tau_{02}$ of the second reagent (see Table 3).

Let us begin from considering the mutual behavior of $\sigma_{th}$ and $\sigma_{sph}$. For all reactions, it is the same: below $B_{0sph}$, the CS accounting for ZPO exceeds $\sigma_{sph}$ whereas above the barrier $\sigma_{sph} = \sigma_{th}$. This behavior qualitatively is not different from the original work [12]. Within the coupled-channels scheme, above the barrier, the influence of the channels is known to be minimal which approximately corresponds to (in our notations) $\sigma_{sph} = \sigma_{th}$.

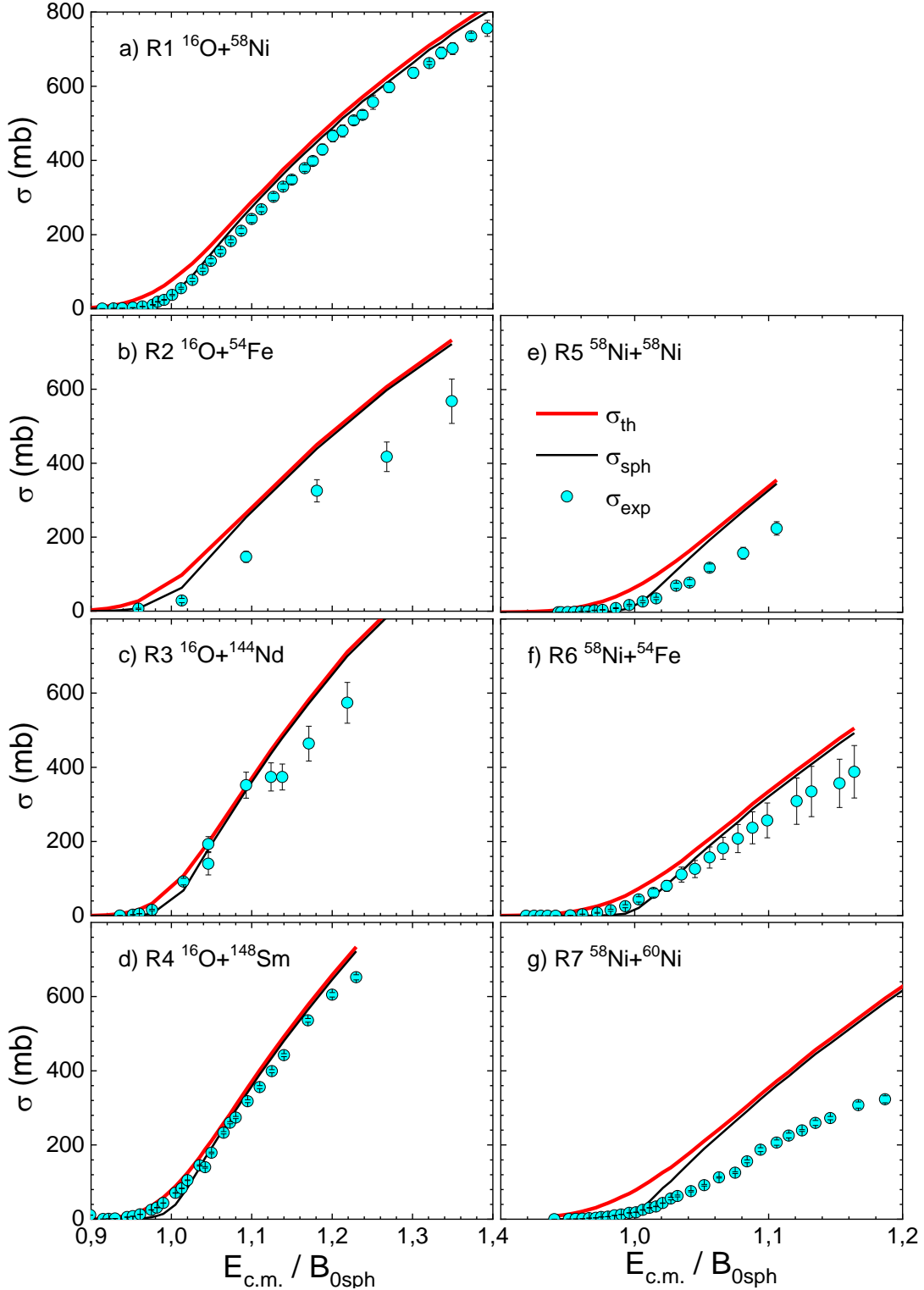

**Fig. 6.** Cross sections versus $E_{c.m.}/B_{0sph}$ in linear scale for reactions R1-R7 (see Table 3). Experimental data from (R1 [55], R2 [56], R3 [57,58], R4 [51], R5 [59], R6 [60], R7 [61]).

Concerning the comparison with the data, at $E_{c.m.} > B_{0sph}$, for all reactions the calculated CSs overshoot the data, as we expected, leaving room for dissipation [32, 34, 35, 71, 72] (not accounted for in the present work). For reactions R1 and R4, this excess is minimal but still presents. However, simultaneous accounting for dissipation and ZPO is a challenging problem [73].

Note that the concept of dissipation in nucleus-nucleus collisions, i.e. conversion of the kinetic energy of relative motion into the intrinsic excitation of the colliding nuclei had been discussed in detail in many works (see, e.g., [64, 74-79]).

Now we go over to the discussion of the relation between the calculated and experimental CSs below the barrier (see Figs. 7). Here, this relation is more involved. For reactions R1-R4 one clearly observes that, for smaller values of $\tau_{02}$, $\sigma_{sph}$ is in better agreement with the data. This corresponds to a sort of averaging of the zero-point deformations as the collision partners approach each other. With the increase of $\tau_{02}$ we see that $\sigma_{th}$ agree better with the data. One could expect something of this sort since with larger $\tau_0$ the deformations have a better chance to be frozen. For reactions R5-R7 in Figs. 7e-g, qualitatively the same kind of agreement between calculated and experimental CSs is seen.

One could argue that the overestimation of the data by $\sigma_{th}$ at the sub-barrier energies for reactions R1, R2, R5-R7 could be compensated by the dissipation as we discussed for the above barrier energies. However, the deeper under the barrier we go, the smaller the role of dissipation must be because the outer turning point is reached at smaller absolute values of the SnnP.

Our present approach to certain extent is similar to that of Ref. [80] which is in turn an extension of the zero-point approach of [12]. In Ref. [80], classical non-dissipative Hamilton equations were solved for the relative motion and quadrupole and octupole deformation parameters of colliding nuclei. The zero-point oscillations enter this approach via the initial conditions for deformation parameters and conjugate momenta. The colliding nuclei were considered to have different relative orientations. At this point, our present approach is more modest. However, in Ref. [80] the SnnP was taken in linear approximation with respect to the deformation parameters and the Coulomb interaction energy was taken in spherical approximation. Moreover, the obsolete Reid parameters of the M3Y NN forces were used in Ref. [80] (see [19]). In our present approach, we calculate both the SnnP and Coulomb interaction exactly (without linear approximation) and employ more actual Paris M3Y NN forces of Ref. [19]. Also in our approach, the quantum tunneling is accounted for, which is ignored in Ref. [80] Concerning our nose-to-nose approximation, we can argue that this geometry results in maximal CSs as demonstrated for instance in Fig. 3 of Ref. [80]. To finalize the comparison between Ref. [80] and present work, we consider seven reactions finding some regularities with respect to the period of quadrupole oscillations of the reagents whereas only one reaction, $^{64}$Ni + $^{64}$Ni, was studied in Ref. [80].

## 6. Conclusions

Basing on the idea of Ref. [12], we have developed a model for evaluating the capture cross sections accounting for the zero-point oscillations (ZPO) of the shape of colliding spherical nuclei. Only quadrupole deformations are accounted for. In contrast to [12], the nucleus-nucleus potential has been evaluated using the semi-microscopic double folding model. The nucleon densities required for these calculations have been generated using the experimental charge densities available in the literature. For the effective nucleon-nucleon forces, the Paris M3Y forces with zero-range exchange part of modified strength have been used. We would like to stress that no fitting parameters are employed in our approach.

The resulting cross sections have been shown to converge as both the number of deformations and the value of the largest accounted deformation increase. The cross sections calculated accounting for ZPO in full power, $\sigma_{th}$, and ignoring these oscillations completely, $\sigma_{sph}$, have been compared to each other and to the experimental data at the above- and below-barrier collision energies $E_{c.m.}$ for seven reactions with the barrier energies ranging from 30 MeV up to 160 MeV.

The mutual behavior of $\sigma_{th}$ and $\sigma_{sph}$ for all reactions is the same: below $B_{0sph}$, the cross section accounting for ZPO exceeds $\sigma_{sph}$ whereas above the barrier $\sigma_{sph} = \sigma_{th}$. This is in agreement with the results of the original work [12] and of the coupled-channels scheme.

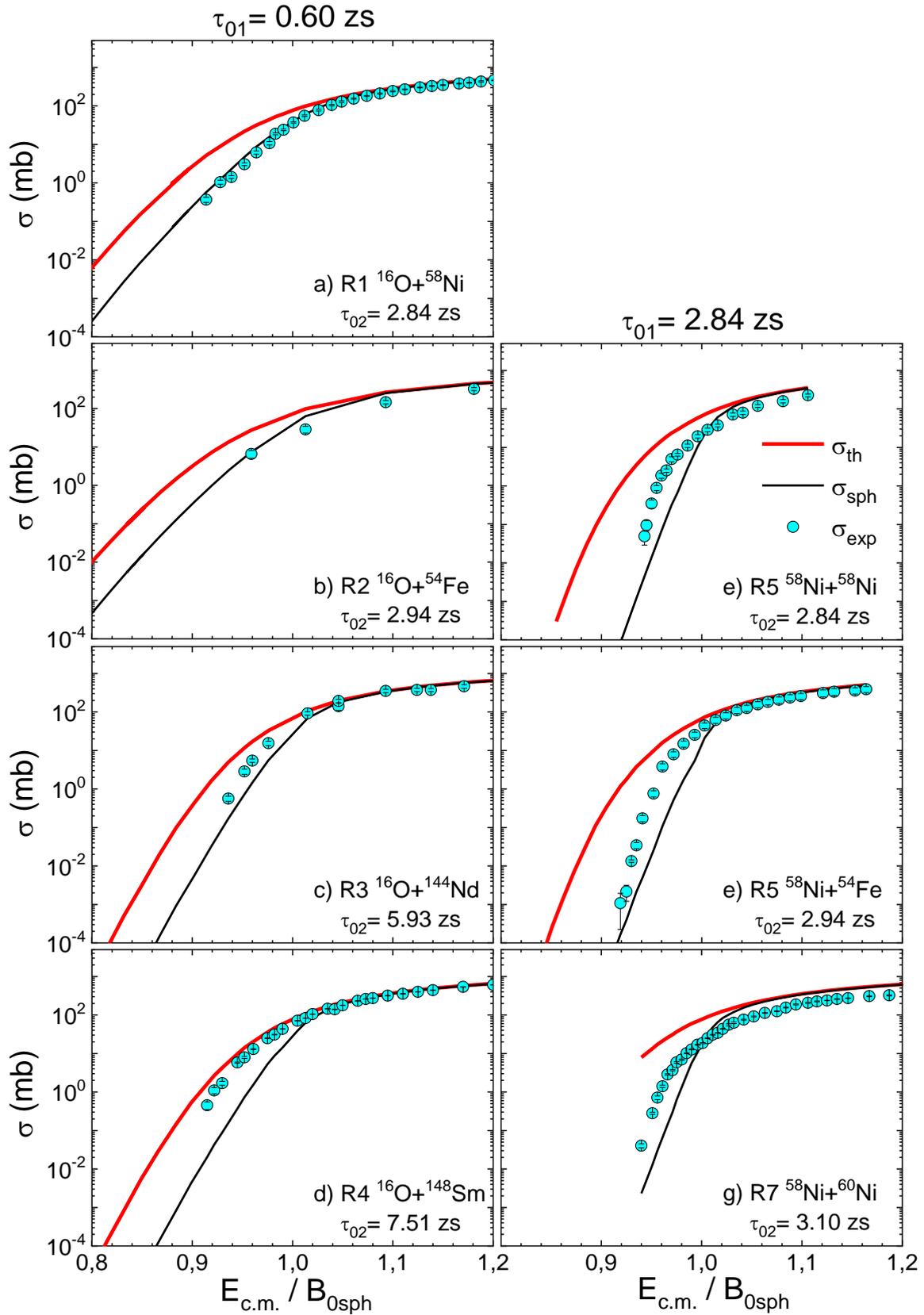

**Fig. 7.** Same as in Fig. 6 but in logarithmic scale. Periods of the second reagent quadrupole oscillations $\tau_{02}$ are indicated in the panels. $\tau_{01}$ denotes the period of the first reagent quadrupole oscillations.

At the above-barrier collision energies, i.e. at $E_{c.m.} > B_{0sph}$, the calculated cross sections exceed the data for all reactions. In our opinion, this is exactly what is needed to have room for including the effect of dissipation ignored in the present work. Accounting for dissipation resulted in an agreement with the above barrier data in Refs. [32-37]. Even those who use and develop the coupled-channels approach for years make efforts to include dissipation in the quantum picture of the barrier penetration model [73, 81, 82]. Moreover, strong dissipation suddenly appearing inside the barrier is implied by the incoming wave boundary conditions which are standard within the coupled-channels approach. Of course, in a more realistic picture, dissipation should rise gradually as the colliding nuclei approach each other [73, 83, 84].

Below the barrier, for all reactions $\sigma_{th} \geq \sigma_{exp} \geq \sigma_{sph}$ that is also reasonable. We have tried analyzing to what extent $\sigma_{th}$ or $\sigma_{sph}$ is closer to $\sigma_{exp}$ considering the period of zero-point oscillations $\tau_0$. Thus, the nuclear structure effects enter our analysis via the rms of ZPO deformation and $\tau_0$. This analysis has enabled us to explain qualitatively the better or worse agreement of $\sigma_{sph}$ and $\sigma_{th}$ with $\sigma_{exp}$.

We plan to upgrade our model in the future by accounting for i) octupole zero-point oscillations, ii) different orientations of the deformed colliding nuclei, iii) dissipation and ZPO simultaneously. At the time being, we consider the latter as a challenging task.

## Acknowledgements

The work was supported by the Foundation for the Advancement of Theoretical Physics and Mathematics "BASIS".

## References

[1]    A. C. Berriman, D. J. Hinde, M. Dasgupta, C. R. Morton, R. D. Butt, J. O. Newton, Unexpected inhibition of fusion in nucleus–nucleus collisions. Nature, 413 (2001) 144–147. doi:10.1038/35093069.
[2]    W. Loveland, An experimentalist's view of the uncertainties in understanding heavy element synthesis, Eur. Phys. J. A. 51 (2015) 120. doi:10.1140/epja/i2015-15120-2.
[3]    B. Back, T. Esbensen, Recent developments in heavy-ion fusion reactions, Rev. Mod. Phys. 86 (2014) 317-360. doi: 10.1103/RevModPhys.86.317.
[4]    V. I. Zagrebaev, Y. Aritomo, M. G. Itkis, Yu. Ts. Oganessian, M. Ohta, Synthesis of superheavy nuclei: How accurately can we describe it and calculate the cross sections, Phys. Rev. C 65 (2001) 014607. doi:10.1103/PhysRevC.65.014607.
[5]    V. V. Sargsyan, Z. Kanokov, G. G. Adamian, N. V. Antonenko, Quantum statistical effects in nuclear reactions, fission, and open quantum systems, Phys. Part. Nucl. 41 (2010) 327-433
[6]    R. Bock, Y. T. Chu, M. Dakowski, A. Gobbi, E. Grosse, A. Olmi, H. Sann, D. Schwalm, U. Lynen, W. Müller, S. Bjornholm, H. Esbensen, W. Wölfli, E. Morenzoni, Dynamics of the fusion process, Nucl. Phys. A 388 (1982) 334-380. doi: 10.1016/0375-9474(82)90420-1.
[7]    C. Gregoire, C. Ngo, E. Tomasi, B. Remaud, F. Scheuter, Fast fission phenomenon. Nucl. Phys. A 387 (1982) 37c-50c. doi: 10.1016/0375-9474(82)90190-7.
[8]    W. J. Swiatecki, Macroscopic treatment of nuclear dynamics, Nucl. Phys. A 428 (1984) 199-222. doi: 10.1016/0375-9474(84)90252-5
[9]    J. P. Blocki, H. Feldmeier, W. J. Swiatecki, Dynamical hindrance to compound-nucleus formation in heavy-ion reactions, Nucl. Phys. A 459 (1986) 145-172. 10.1016/0375-9474(86)90061-8
[10]   Sh. A. Kalandarov, D. Lacroix, G. G. Adamian, N. V. Antonenko, J. P. Wieleczko, S. Pirrone, G. Politi, Quasifission and fusion-fission processes in the reactions 78Kr + 40Ca and 86Kr + 48Ca at 10 MeV/nucleon bombarding energy, Phys. Rev. C 93 (2016) 024613. doi:10.1103/PhysRevC.93. 024613.
[11]   Kyle Goodbey, A. S. Umar, Quasifission dynamics in microscopic theories, Front. Phys. 8 (2020) 40. doi: 10.3389/fphy.2020.00040.
[12]   H. Esbensen, Fusion and zero-point motions, Nucl. Phys. A. 352 (1981) 147–156. doi:10.1016/0375-9474(81)90565-0.
[13]   S. Landowne, J.R. Nix, Two-dimensional calculation of sub-barrier heavy-ion fusion cross sections, Nucl. Phys. A. 368 (1981) 352–364. doi:10.1016/0375-9474(81)90690-4.
[14]   W. Reisdorf, F.P. Hessberger, K.D. Hildenbrand, S. Hofmann, G. Münzenberg, K.-H. Schmidt, J.H.R. Schneider, W.F.W. Schneider, K. Sümmerer, G. Wirth, J. V. Kratz, K. Schlitt, Influence of Collective Surface Motion on the Threshold Behavior of Nuclear Fusion, Phys. Rev. Lett. 49 (1982) 1811–1815. doi:10.1103/PhysRevLett.49.1811.
[15]   G. Bertsch, J. Borysowicz, H. McManus, W.G. Love, Interactions for inelastic scattering derived from realistic potentials, Nucl. Phys. A. 284 (1977) 399–419. doi:10.1016/0375-9474(77)90392-X.

[16] M. Ismail, K.A. Ramadan, Microscopic calculation of sub-barrier fusion cross section and barrier distribution using M3Y-type forces, J. Phys. G Nucl. Part. Phys. 26 (2000) 1621–1633. doi:10.1088/0954-3899/26/10/312.
[17] I.I. Gontchar, D.J. Hinde, M. Dasgupta, J.O. Newton, Double folding nucleus-nucleus potential applied to heavy-ion fusion reactions, Phys. Rev. C. 69 (2004) 024610. doi:10.1103/PhysRevC.69.024610.
[18] M. Bhuyan, R. Kumar, Fusion cross section for Ni-based reactions within the relativistic mean-field formalism, Phys. Rev. C. 98 (2018) 054610. doi:10.1103/PhysRevC.98.054610.
[19] N. Anantaraman, H. Toki, G.F. Bertsch, An effective interaction for inelastic scattering derived from the Paris potential, Nucl. Phys. A. 398 (1983) 269–278. doi:10.1016/0375-9474(83)90487-6.
[20] M. Dasgupta, D.J. Hinde, N. Rowley, A.M. Stefanini, Measuring barriers to fusion, Annu. Rev. Nucl. Part. Sci. 48 (1998) 401–461. doi:10.1146/annurev.nucl.48.1.401
[21] J.O. Newton, C.R. Morton, M. Dasgupta, J.R. Leigh, J.C. Mein, D.J. Hinde, H. Timmers, K. Hagino, Experimental barrier distributions for the fusion of 12C, 16O, 28Si, and 35Cl with 92Zr and coupled-channels analyses, Phys. Rev. C. 64 (2001) 064608. doi:10.1103/PhysRevC.64.064608.
[22] A. Mukherjee, D.J. Hinde, M. Dasgupta, K. Hagino, J.O. Newton, R.D. Butt, Failure of the Woods-Saxon nuclear potential to simultaneously reproduce precise fusion and elastic scattering measurements, Phys. Rev. C. 75 (2007) 044608. doi:10.1103/PhysRevC.75.044608.
[23] K.T. Lesko, W. Henning, K.E. Rehm, G. Rosner, J.P. Schiffer, G.S.F. Stephans, B. Zeidman, W.S. Freeman, Fission following fusion of Ni + Sn, Phys. Rev. C. 34 (1986) 2155–2164. doi:10.1103/PhysRevC.34.2155.
[24] L. Jiang, A.M. Stefanini, H. Esbensen, K.E. Rehm, S. Almaraz-Calderon, M.L. Avila, B.B. Back, D. Bourgin, L. Corradi, S. Courtin, E. Fioretto, F. Galtarossa, A. Goasduff, F. Haas, M.M. Mazzocco, D. Montanari, G. Montagnoli, T. Mijatovic, R. Sagaidak, D. Santiago-Gonzalez, F. Scarlassara, E.E. Strano, S. Szilner, Fusion reactions of 58,64Ni+124Sn, Phys. Rev. C. 91 (2015) 044602. doi:10.1103/PhysRevC.91.044602.
[25] W. Reisdorf, F.P. Hessberger, K.D. Hildenbrand, S. Hofmann, G. Münzenberg, K.H. Schmidt, J.H.R. Schneider, W.F.W. Schneider, K. Sümmerer, G. Wirth, J. V. Kratz, K. Schlitt, Fusion near the threshold: A comparative study of the systems 40Ar + 112, 116, 122sn and 40Ar + 144, 148, 154Sm, Nucl. Phys. A. 438 (1985) 212–252. doi:10.1016/0375-9474(85)90125-3.
[26] R. A. Broglia, C. H. Dasso, S. Landowne, G. Pollarolo, Estimate of enhancements in sub-barrier heavy-ion fusion cross sections due to coupling to inelastic and transfer reaction channels, Phys. Lett. B 133 (1983) 34-38. doi: 10.1016/0370-2693(83)90100-4.
[27] M. Beckerman // Sub-barrier fusion of two nuclei // Rep. Prog. Phys. 51 (1988) 1047-1103.
[28] K. Hagino, N. Rowley, A.T. Kruppa, A program for coupled-channel calculations with all order couplings for heavy-ion fusion reactions, Comput. Phys. Commun. 123 (1999) 143–152. doi:https://doi.org/10.1016/S0010-4655(99)00243-X.
[29] V.I. Zagrebaev, V. V Samarin, Near-barrier fusion of heavy nuclei: Coupling of channels, Phys. At. Nucl. 67 (2004) 1462–1477. doi:10.1134/1.1788037.
[30] J.O. Newton, R.D. Butt, M. Dasgupta, D.J. Hinde, I.I. Gontchar, C.R. Morton, K. Hagino, Systematic failure of the Woods-Saxon nuclear potential to describe both fusion and elastic scattering: Possible need for a new dynamical approach to fusion, Phys. Rev. C. 70 (2004) 024605. doi:10.1103/PhysRevC.70.024605.
[31] J.O. Newton, R.D. Butt, M. Dasgupta, D.J. Hinde, I.I. Gontchar, C.R. Morton, K. Hagino, Systematics of precise nuclear fusion cross sections: The need for a new dynamical treatment of fusion?, Phys. Lett. B. 586 (2004) 219–224. doi:10.1016/j.physletb.2004.02.052.
[32] P. Fröbrich, Fusion and capture of heavy ions above the barrier: Analysis of experimental data with the surface friction model, Phys. Rep. 116 (1984) 337–400. doi:10.1016/0370-1573(84)90162-5.
[33] P. Fröbrich, I.I. Gontchar, Langevin description of fusion, deep-inelastic collisions and heavy-ion-induced fission, Phys. Rep. 292 (1998) 131–237. doi:10.1016/S0370-1573(97)00042-2.
[34] I.I. Gontchar, R. Bhattacharya, M. V. Chushnyakova, Quantitative analysis of precise heavy-ion fusion data at above-barrier energies using Skyrme-Hartree-Fock nuclear densities, Phys. Rev. C. 89 (2014) 034601. doi:10.1103/PhysRevC.89.034601.
[35] M. V. Chushnyakova, R. Bhattacharya, I.I. Gontchar, Dynamical calculations of the above-barrier heavy-ion fusion cross sections using Hartree-Fock nuclear densities with the SKX coefficient set, Phys. Rev. C. 90 (2014) 017603. doi:10.1103/PhysRevC.90.017603.

[36] R.A. Kuzyakin, V. V. Sargsyan, G.G. Adamian, N. V. Antonenko, Quantum Diffusion Description of Large-Amplitude Collective Nuclear Motion, Phys. Part. Nucl. 48 (2017) 21–118

[37] V.V. Sargsyan, S.Y. Grigoryev, G.G. Adamian, N.V. Antonenko, Capture cross section with quantum diffusion approach, Comput. Phys. Commun. 233 (2018) 145–155. doi:10.1016/J.CPC.2018.06.011.

[38] R. Van den Bossche, A. Diaz-Torres, Production of transuranium isotopes in 20Ne-induced incomplete fusion reactions, Phys. Rev. C. 102 (2020) 064618. doi:10.1103/PhysRevC.102.064618.

[39] V.L. Litnevsky, F.A. Ivanyuk, G.I. Kosenko, S. Chiba, Formation of superheavy nuclei in 36S+238U and 64Ni+238U reactions, Phys. Rev. C. 101 (2020) 064616. doi:10.1103/PhysRevC.101.064616.

[40] G.R. Satchler, W.G. Love, Folding model potentials from realistic interactions for heavy-ion scattering, Phys. Rep. 55 (1979) 183–254. doi:10.1016/0370-1573(79)90081-4.

[41] Dao T. Khoa, W. von Oertzen, H. G. Bohlen, Double-folding model for heavy-ion optical potential: Revised and applied to study 12C and 16O elastic scattering, Phys. Rev. C 49 (1994) 1652-1668. doi: 10.1103/PhysRevC.49.1652

[42] L. H. Chien, D. T. Khoa, D. C. Cuong, N. H. Phuc, Consistent mean-field description of the 12C+12C optical potential at low energies and the astrophysical S factor, Phys. Rev. C 98 (2018) 064604. doi:10.1103/PhysRevC.98.064604.

[43] D.T. Khoa, G.R. Satchler, W. von Oertzen, Nuclear incompressibility and density dependent NN interactions in the folding model for nucleus-nucleus potentials, Phys. Rev. C. 56 (1997) 954–969. doi:10.1103/PhysRevC.56.954.

[44] D.T. Khoa, O.M. Knyazkov, Exchange effects in elastic and inelastic alpha- and heavy-ion scattering, Zeitschrift Für Phys. A At. Nucl. 328 (1987) 67–79. doi:10.1007/BF01295184.

[45] M. V. Chushnyakova, I.I. Gontchar, O.M. Sukhareva, N.A. Khmyrova, Modification of the effective Yukawa-type nucleon–nucleon interaction for accelerating calculations of the real part of the optical potential, Moscow Univ. Phys. Bull. 76 (2021) 202–208. doi:10.3103/S0027134921040056.

[46] I.I. Gontchar, M. V. Chushnyakova, O.M. Sukhareva, Systematic application of the M3Y NN forces for describing the capture process in heavy-ion collisions involving deformed target nuclei, Phys. Rev. C. 105 (2022) 014612. doi:10.1103/PhysRevC.105.014612.

[47] M.J. Rhoades-Brown, V.E. Oberacker, M. Seiwert, W. Greiner, Potential pockets in the 238U+238U system and their possible consequences, Zeitschrift Für Phys. A. 310 (1983) 287–294. doi:10.1007/BF01419514.

[48] M. V. Chushnyakova, I.I. Gontchar, Post-scission dissipative motion and fission-fragment kinetic energy, Bull. Russ. Acad. Sci. Phys. 80 (2016) 938–941. doi:10.3103/S1062873816080086.

[49] H. De Vries, C.W. De Jager, C. De Vries, Nuclear charge-density-distribution parameters from elastic electron scattering, At. Data Nucl. Data Tables. 36 (1987) 495–536. doi:10.1016/0092-640X(87)90013-1.

[50] I.I. Gontchar, D.J. Hinde, M. Dasgupta, C.R. Morton, J.O. Newton, Semi-microscopic calculations of the fusion barrier distributions for reactions involving deformed target nuclei, Phys. Rev. C. 73 (2006) 034610. doi:10.1103/PhysRevC.73.034610.

[51] J.R. Leigh, M. Dasgupta, D.J. Hinde, J.C. Mein, C.R. Morton, R.C. Lemmon, J.P. Lestone, J.O. Newton, H. Timmers, J.X. Wei, N. Rowley, Barrier distributions from the fusion of oxygen ions with 144,148,154Sm and 186W, Phys. Rev. C. 52 (1995) 3151–3166. doi:10.1103/PhysRevC.52.3151.

[52] A.V. Karpov, A.S. Denikin, M.A. Naumenko, A.P. Alekseev, V.A. Rachkov, V.V. Samarin, V.V. Saiko, V.I. Zagrebaev, NRV web knowledge base on low-energy nuclear physics, Nucl. Instruments Methods Phys. Res. Sect. A Accel. Spectrometers, Detect. Assoc. Equip. 859 (2017) 112–124 doi:10.1016/j.nima.2017.01.069. see also http://nrv.jinr.ru/nrv/

[53] H. Esbensen, S. Landowne, Higher-order coupling effects in low energy heavy-ion fusion reactions, Phys. Rev. C. 35 (1987) 2090–2096. doi:10.1103/PhysRevC.35.2090.

[54] K. Hagino, N. Takigawa, M. Dasgupta, D.J. Hinde, J.R. Leigh, Validity of the linear coupling approximation in heavy-ion fusion reactions at sub-barrier energies, Phys. Rev. C. 55 (1997) 276–284. doi:10.1103/PhysRevC.55.276.

[55] N. Keeley, J.S. Lilley, J.X. Wei, M. Dasgupta, D.J. Hinde, J.R. Leigh, J.C. Mein, C.R. Morton, H. Timmers, N. Rowley, Fusion excitation function measurements for the 16O+58Ni and 16O+62Ni systems, Nucl. Phys. A. 628 (1998) 1–16. doi:10.1016/S0375-9474(97)00597-6.

[56] H. Funaki, E. Arai, Anomaly in the 15N, 16O, 19F + 54,56Fe fusion cross sections around the Coulomb barrier energy, Nucl. Phys. A. 556 (1993) 307–316. doi:10.1016/0375-9474(93)90353-Y.

[57] G. Duchêne, P. Romain, F.A. Beck, P. Benet, D. Disdier, B. Haas, B. Lott, V. Rauch, F. Scheibling, J.P. Vivien, S.K. Basu, E. Bozek, K. Zuber, D. Di Gregorio, J. Fernandez-Niello, Angular momentum distributions for 16O+144Nd, Phys. Rev. C. 47 (1993) 2043–2054. doi:10.1103/PhysRevC.47.2043.

[58] M. di Tada, D.E. DiGregorio, D. Abriola, O.A. Capurro, G. Duchêne, M. Elgue, A. Etchegoyen, J.O. Fernández Niello, A.M.J. Ferrero, A.J. Pacheco, P.R. Silveira Gomes, J.E. Testoni, Sub-barrier fusion of 16O+144Nd, Phys. Rev. C. 47 (1993) 2970–2973. doi:10.1103/PhysRevC.47.2970.

[59] M. Beckerman, J. Ball, H. Enge, M. Salomaa, A. Sperduto, S. Gazes, A. DiRienzo, J.D. Molitoris, Near- and sub-barrier fusion of 58Ni with 58Ni, Phys. Rev. C. 23 (1981) 1581–1589. doi:10.1103/PhysRevC.23.1581.

[60] A.M. Stefanini, G. Montagnoli, L. Corradi, S. Courtin, D. Bourgin, E. Fioretto, A. Goasduff, J. Grebosz, F. Haas, M. Mazzocco, T. Mijatović, D. Montanari, M. Pagliaroli, C. Parascandolo, F. Scarlassara, E. Strano, S. Szilner, N. Toniolo, D. Torresi, Fusion of 48Ti+58Fe and 58Ni+54Fe below the Coulomb barrier, Phys. Rev. C. 92 (2015) 064607. doi:10.1103/PhysRevC.92.064607.

[61] A.M. Stefanini, D. Ackermann, L. Corradi, D.R. Napoli, C. Petrache, P. Spolaore, P. Bednarczyk, H.Q. Zhang, S. Beghini, G. Montagnoli, L. Mueller, F. Scarlassara, G.F. Segato, F. Soramel, N. Rowley, Influence of Complex Surface Vibrations on the Fusion of 58Ni+60Ni, Phys. Rev. Lett. 74 (1995) 864–867. doi:10.1103/PhysRevLett.74.864.

[62] M. Beckerman, Sub-barrier fusion of atomic nuclei, Phys. Rep. 129 (1985) 145-223 doi: 10.1016/0370-1573(85)90058-4.

[63] G.G. Adamian, N.V. Antonenko, R. V Jolos, S.P. Ivanova, O.I. Melnikova, Effective nucleus-nucleus potential for calculation of potential energy of a dinuclear system, Int. J. Mod. Phys. E. 5 (1996) 191–216. doi:10.1142/S0218301396000098.

[64] P. Fröbrich and R. Lipperheide, *Theory of Nuclear Reactions*, Oxford Studies in Nuclear Physics Vol. 18 (Oxford University, Oxford, 1996). p.281

[65] R. G. Stokstad, Y. Eisen, S. Kaplanis, D. Pelte, U. Smilansky, I. Tserruya, Effect of Nuclear Deformation on Heavy-Ion Fusion, Phys. Rev. Lett. 41 (1978) 465-469. doi:10.1104/PhysRevLett.41.465

[66] R. G. Stokstad, E. E. Gross, Analysis of the sub-barrier fusion of 16O+148, 150, 152, 154 Sm, Phys. Rev. C 23 (1978) 281-294. doi:10.1104/PhysRevC.23.281.

[67] S.A. Alavi, V. Dehghani, Investigation of heavy-ion fusion with deformed surface diffuseness: Actinide and lanthanide targets, Phys. Rev. C. 95 (2017) 054602. doi:10.1103/PhysRevC.95.054602.

[68] M. R. Pahlavani, N. Karamzadeh, Partial α-decay half-lives of ground to ground and ground to excited states of Thorium family, Chinese J. Phys. 56 (2018) 1727–1733. doi:10.1016/J.CJPH.2018.05.014.

[69] J. J. Bekx, M. L. Lindsey, S. H. Glenzer, K.-G. Schlesinger, Applicability of semiclassical methods for modeling laser-enhanced fusion rates in a realistic setting, Phys. Rev. C. 105 (2022) 054001. doi:10.1103/PhysRevC.105.054001.

[70] R.A. Gherghescu, D.N. Poenaru, Fission channels for fragment isotopes from 298Fl with magic nucleon numbers, Phys. Rev. C. 106 (2022) 034611. doi:10.1103/PhysRevC.106.034611.

[71] M. V. Chushnyakova, I. I. Gontchar, Heavy ion fusion: Possible dynamical solution of the problem of the abnormally large diffuseness of the nucleus-nucleus potential, Phys. Rev. C. 87 (2013) 014614. doi:10.1103/PhysRevC.87.014614.

[72] D. Y. Jeung, D. J. Hinde, E. Williams, M. Dasgupta, E. C. Simpson, R. du Rietz, D. H. Luong, R. Rafiei, M. Evers, I. P. Carter, K. Ramachandran, C. Palshetkar, D. C. Rafferty, C. Simenel, A. Wakhle, Energy dissipation and suppression of capture cross sections in heavy ion reactions, Phys. Rev. C. 103 (2021) 034603. doi:10.1103/PhysRevC.103.034603.

[73] E. Piasecki, M. Kowalczyk, S. Yusa, A. Trzcinska, K. Hagino, Dissipation and tunneling in heavy-ion reactions near the Coulomb barrier, Phys. Rev. C 100 (2019) 014616. doi: 10.1103/PhysRevC.100. 014616.

[74] D. H. E. Gross, H. Kalinowski, Friction model of heavy-ion collisions, Phys. Rep. 45 (1978) 175-210. doi:10.1016/0370-1573(78)90031-5.

[75] V. V. Volkov, Deep inelastic transfer reactions — The new type of reactions between complex nuclei, Phys. Rep. 44 (1978) 93-157. doi: 10.1016/0370-1573(78)90200-4.

[76] R. Beck, D. H. E. Gross, Nuclear friction in heavy ion scattering, Phys. Lett. B 47 (1973) 143-146. doi: 10.1016/0370-2693(73)90591-1.

[77] W. J. Swiatecki, The Dynamics of Nuclear Coalescence or Reseparation, Phys. Scr. 24 (981) 113-122. doi: 10.1088/0031-8949/24/1B/007.

[78] G.G. Adamian, N.V. Antonenko, W. Scheid, Friction and diffusion coefficients in coordinate in nonequilibrium nuclear processes, Nucl. Phys. A 645 (1999) 376-398. doi: 10.1016/S0375-9474(98)00560-0.

[79] P. Fröbrich, Dynamical calculation of fusion and capture cross sections in heavy-ion collisions, Phys. Lett. B 122 (1983) 338-342. doi: 10.1016/0370-2693(83)91577-0.

[80] S. Ayik, B. Yilmaz, D. Lacroix, Stochastic semi-classical description of fusion at near-barrier energies, Phys. Rev. C. 81 (2010) 034605. doi:10.1103/PhysRevC.81.034605.

[81] S. Yusa, K. Hagino, N. Rowley, Quasi-elastic scattering in the 20Ne+90,92Zr reactions: Role of noncollective excitations, Phys. Rev. C. 88 (2013) 054621. doi:10.1103/PhysRevC.88.054621.

[82] M. Tokieda, K. Hagino, Quantum tunneling with friction, Phys. Rev. C. 95 (2017) 054604. doi:10.1103/PhysRevC.95.054604.

[83] M. Dasgupta, D.J. Hinde, A. Diaz-Torres, B. Bouriquet, C.I. Low, G.J. Milburn, J.O. Newton, Beyond the Coherent Coupled Channels Description of Nuclear Fusion, Phys. Rev. Lett. 99 (2007) 192701. doi:10.1103/PhysRevLett.99.192701.

[84] A. Diaz-Torres, D.J. Hinde, M. Dasgupta, G.J. Milburn, J.A. Tostevin, Dissipative quantum dynamics in low-energy collisions of complex nuclei, Phys. Rev. C. 78 (2008) 064604. doi:10.1103/PhysRevC.78.064604.